\documentclass[prd,preprint,tightenlines,floatfix,showpacs,preprintnumbers,nofootinbib,eqsecnum]{revtex4}

 \usepackage[dvips,final]{graphicx}
  \usepackage{amssymb}
   \usepackage{amsmath}
    \usepackage{amsfonts}
     \usepackage{epsfig}
      \usepackage{bm}% bold math
\begin{document}

\thispagestyle{empty} \preprint{\hbox{}} \vspace*{-10mm}

\title{Diffractive production
of $\chi_c(0^+,1^+,2^+)$ mesons\\ at LHC, Tevatron and
RHIC\footnote{Talk given at the Europhysics Conference on High
Energy Physics, EPS-HEP 2009, July 16 - 22 2009, Krakow, Poland}}

\author{R.~S.~Pasechnik}
\email{rpasech@theor.jinr.ru}

\author{A.~Szczurek}
\email{antoni.szczurek@ifj.edu.pl}

\author{O.~V.~Teryaev}
\email{teryaev@theor.jinr.ru}

\affiliation{ Bogoliubov Laboratory of Theoretical Physics, JINR,
Dubna 141980, Russia}

\affiliation{Institute of Nuclear Physics PAN, PL-31-342 Cracow,
Poland} \affiliation{University of Rzesz\'ow, PL-35-959 Rzesz\'ow,
Poland}

\begin{abstract}
We consider exclusive double diffractive production of scalar
$\chi_c(0^+)$, axial-vector $\chi_c(1^+)$ and tensor $\chi_c(2^+)$
charmonia in proton-(anti)proton collisions at different energies.
The corresponding amplitudes for these processes are derived within
the $k_t$-factorisation approach and the corresponding cross section
is calculated with different unintegrated gluon distribution
functions (UGDFs) known from the literature. We compare exclusive
production of all charmonium states $\chi_c(0^+),\,\chi_c(1^+)$ and
$\chi_c(2^+)$ including branching fraction for radiative $J/\Psi +
\gamma$ decay channel. Corresponding experimental consequences are
discussed.
\end{abstract}

%Keywords:

\maketitle

\begin{figure}[!h]    % Figure 1
 \centerline{\includegraphics[width=0.23\textwidth]{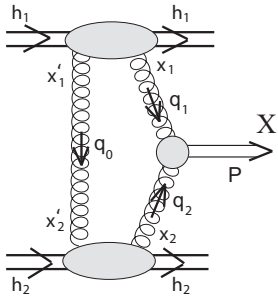}}
   \caption{\label{fig:fig1}
   \small  The QCD mechanism of diffractive production of the heavy central system $X$.}
\end{figure}

The QCD mechanism for the diffractive production of heavy central
system has been proposed recently by Kaidalov, Khoze, Martin and
Ryskin (Durham group, KKMR) for Higgs production at the LHC (see
Refs.~\cite{KMR,KKMR,KKMR-spin}). In the framework of this approach
the amplitude of the exclusive $pp\to pXp$ process is considered to
be a convolution of the hard subprocess amplitude describing the
fusion of two off-shell gluons into a heavy system $g^*g^*\to X$,
and the soft part represented in terms of the off-diagonal
unintegrated gluon distributions (UGDFs) (see Fig.~\ref{fig:fig1}).
%--------------------------------------------------------------------
%--------------------------------------------------------------------

In order to check the underlying production mechanism it is worth to
replace Higgs boson by a lighter (but still heavy enough to provide
the factorisation) meson which is easier to measure. In this respect
the exclusive production of heavy quarkonia is under special
interest from both experimental and theoretical points of view
\cite{exp-thr}. Testing the KKMR approach against various data on
exclusive meson production at high energies is a good probe of
nonperturbative dynamics of partons described by UGDFs.

Recently, the signal from the diffractive $\chi_c(0^+,1^+,2^+)$
charmonia production in the radiative $J/\Psi + \gamma$ decay
channel has been measured by the CDF Collaboration
\cite{Aaltonen:2009kg}: $d\sigma/dy|_{y=0}(pp\to
pp(J/\psi+\gamma))\simeq0.97\pm0.26$. In the very forward limit the
contributions from $\chi_c(1^+,2^+)$ vanish due to the $J_z=0$
selection rule, however, for general kinematics this might not be
true \cite{Martin:2009ku}. In particular, it was shown in
Ref.~\cite{Pasechnik:2009bq} that the axial-vector $\chi_c(1^+)$
production, due a relatively large branching fraction of its
radiative decay, may not be negligible and gives a noticeable
contribution to the total signal measured by the CDF Collaboration.
As shown below, the same holds also for the tensor $\chi_c(2^+)$
meson contribution.

The production of the axial-vector $\chi_c(1^+)$ meson has an
additional suppression w.r.t. $\chi_c(0^+,2^+)$ in the limit of
on-shell fusing gluons due to the Landau-Yang theorem
\cite{Pasechnik:2009bq}. Such an extra suppression may lead to the
dominance of the $\chi_c(2^+)$ contribution in the radiative decay
channel. Off-shell effects play a significant role also for the
scalar $\chi_c(0^+)$ production reducing the total cross section by
a factor of 2 -- 5 depending on UGDFs \cite{Pasechnik:2007hm}.

According to the KKMR approach the amplitude of the exclusive double
diffractive color singlet production $pp\to pp\chi_{cJ}$ is
\cite{Khoze:2004yb,Pasechnik:2007hm} {\small \begin{eqnarray} {\cal
M}_{J,\lambda}=const\cdot\delta_{c_1c_2}\Im\int d^2
q_{0,t}V^{c_1c_2}_{J,\lambda}(q_1,q_2,P)
\frac{f^{\mathrm{off}}_{g,1}(x_1,x_1',q_{0,t}^2,q_{1,t}^2,t_1)
      f^{\mathrm{off}}_{g,2}(x_2,x_2',q_{0,t}^2,q_{2,t}^2,t_2)}
{q_{0,t}^2\,q_{1,t}^2\, q_{2,t}^2} \; ,\label{ampl}
\end{eqnarray}}
where $t_{1,2}$ are the momentum transfers along the proton lines,
$q_0$ is the momentum of the screening gluon, $q_{1,2}$ are the
momenta of fusing gluons, and
$f^{\mathrm{off}}_{g,i}(x_i,x_i',q_{0,t}^2,q_{i,t}^2,t_i)$ are the
off-diagonal UGDFs. The prescription for the off-diagonal UGDFs
through their diagonal counterparts, inspired by the positivity
constraints \cite{posit}, reads \cite{SPT07,Pasechnik:2007hm}
{\small
\begin{eqnarray}\nonumber f_{g,1}^{\mathrm{off}} &=&
\sqrt{f_{g}^{(1)}(x_1',q_{0,t}^2,\mu_0^2) \cdot
f_{g}^{(1)}(x_1,q_{1,t}^2,\mu^2)} \cdot F_1(t_1)\,, \\
f_{g,2}^{\mathrm{off}} &=& \sqrt{f_{g}^{(2)}(x_2',q_{0,t}^2,\mu_0^2)
\cdot f_{g}^{(2)}(x_2,q_{2,t}^2,\mu^2)} \cdot F_1(t_2)\,,\quad
x'_1=x'_2=\xi\cdot q_{0,t}/\sqrt{s}\,, \label{skewed_UGDFs}
\end{eqnarray} }
where $F_1(t_1)$ and $F_1(t_2)$ are the isoscalar nucleon form
factors and $\xi$ is auxiliary  parameter.

The hard vertex function $V^{c_1c_2}_{J,\lambda}(q_1,q_2,P)$
describes the coupling of two virtual gluons to $\chi_{cJ}$ meson.
It can be found by using the
next-to-leading-logarithmic-approximation (NLLA) BFKL $g^*g^*(q\bar
q)$-vertex in quasi-multi-Regge kinematics (QMRK) and projecting it
out to the colour singlet bound state $\chi_{cJ}$ employing the
pNRQCD technique (for scalar and axial-vector case, see
Refs.~\cite{Pasechnik:2007hm,Pasechnik:2009bq}). We do not take into
account the NLO QCD corrections here, $K_{NLO}=1$.

Results for the total cross section of diffractive
$\chi_c(0^+,1^+,2^+)$ meson production at Tevatron energy $W = 1960$
GeV are shown in Table~\ref{table:tot}. As have been pointed out in
Ref.~\cite{Martin:2009ku} the absorptive corrections are quite
sensitive to the meson spin-parity. This was studied before in the
context of scalar and pseudoscalar Higgs production in
Ref.~\cite{KKMR}. In the last column we adopt the following
effective gap survival factors (for $\langle p_t \rangle\simeq 0.5$
GeV), calculated for different spins in Ref.~\cite{Martin:2009ku}:
$\langle S_{\mathrm{eff}}^2(\chi_c(0^+))\rangle\simeq0.02$, $\langle
S_{\mathrm{eff}}^2(\chi_c(1^+))\rangle\simeq0.05$ and $\langle
S_{\mathrm{eff}}^2(\chi_c(2^+))\rangle\simeq0.05$. Having in mind
that the total cross section $\sigma_{\chi_c}$ is less than the
differential one $d\sigma_{\chi_c}/dy|_{y=0}$ by a factor of 4 -- 5,
we conclude that the calculated signal in the $J/\psi + \gamma$
channel turns out to be quite below the CDF data for off-diagonal
UGDFs calculated as in Eq.~(\ref{skewed_UGDFs}) with $\xi=1$. This
provides an argument that $x'$ should be smaller than used, i.e.
$\xi<1$. The dependence of the results on $\xi$ will be reported
elsewhere.
\begin{table}[!h]
\caption{\label{table:tot} \small Integrated cross section
$\sigma_{\chi_c}$ (in nb) of the exclusive diffractive production of
$\chi_c(0^+,1^+,2^+)$ mesons and their signals in radiative decay
channel for different UGDFs at Tevatron. First three lines
correspond to Eq.~(\ref{skewed_UGDFs}) with $\xi=1$, forth line is
calculated in the KMR prescription \cite{KMR} with $R_g=1$. In the
last column we took into account the NLO corrections with
$K_{NLO}\simeq 1.5$.}
 {\small
\begin{center}
\begin{tabular}{|c||c|c||c|c||c|c||c|c||}
\hline
 &\multicolumn{2}{l||}{$\quad\,\;\chi_c(0^+)$}&
  \multicolumn{2}{l||}{$\quad\,\;\chi_c(1^+)$}&
  \multicolumn{2}{l||}{$\quad\,\;\chi_c(2^+)$}& ratio & expected signal \\
\cline{2-7}
 UGDF &$\sigma_{\chi_c}$&$\sigma_{J/\psi\gamma}$&
 $\sigma_{\chi_c}$&$\sigma_{J/\psi\gamma}$&$\sigma_{\chi_c}$&$\sigma_{J/\psi\gamma}$&
 $\frac{\chi_c(2^+)\to J/\psi\gamma}{\chi_c(0^+)\to J/\psi\gamma}$&$K_{NLO}^2\sum_{\chi_c}
 \langle S_{\mathrm{eff}}^2\rangle\sigma_{J/\psi\gamma}$ \\
\hline\hline
KL   \cite{KL}      & 55.2  & 0.6  & 0.5  &  0.2 & 6.7   & 1.3  & 2.2 & 0.2  \\
GBW  \cite{GBW}     & 160   & 1.8  & 4.2  &  1.4 & 50.2  & 9.7  & 5.4 & 1.3  \\
KS \cite{KS05}      & 443.6 & 5.1  & 3.0  &  1.0 & 50.8  & 9.9  & 1.9 & 1.5  \\
KMR  \cite{KMR}     & 1127  & 12.8 & 8.1  &  2.8 & 74.2  & 14.4 & 1.1 & 2.5  \\
 \hline
\end{tabular}
\end{center}}
\end{table}

The relative contributions of different charmonium states in
$J/\psi+\gamma$ channel are found to be:
\begin{displaymath}
 \sigma(0^+\to J/\psi+\gamma):\sigma(1^+\to
J/\psi+\gamma):\sigma(2^+\to J/\psi+\gamma) = \left\{
     \begin{array}{lr}
       1:0.71:4.64,  \qquad\mathrm{KL}\\
       1:1.94:13.47,   \;\quad\mathrm{GBW}\\
       1:0.49:4.85,   \qquad\mathrm{KS}\\
       1:0.55:2.81,   \qquad\mathrm{KMR}
     \end{array}
   \right.
\end{displaymath}
We see that the contribution of the tensor $\chi_c(2^+)$ meson
dominates over $\chi_c(0^+,1^+)$ for all UGDFs.

It is also interesting to consider the diffractive production of
$\chi_c$ states at different energies. As an example, in
Table~\ref{table:Edep} we present the integrated cross sections of
$\chi_c(1^+)$ production at RHIC, Tevatron and LHC energies
\cite{Pasechnik:2009bq}. The results show similar energy behavior of
the diffractive cross section for different UGDFs except KMR UGDF
giving more strong growth of the cross section with energy. The same
is true for other $\chi_c$ states.
\begin{table}[!h]
\caption{\label{table:Edep} \small Integrated cross section
$\sigma_{tot}$ (in nb) for exclusive $\chi_c(1^+)$ production at
different energies. No absorption effects are included here.}
\begin{center}
\begin{tabular}{|c|c|c|c|}
\hline
UGDF &$\quad$RHIC$\quad$&$\quad$Tevatron$\quad$& $\quad$LHC$\quad$\\
\hline
KL                   & 0.05  & 0.5  & 1.7   \\
GBW                  & 0.04  & 4.2  & 73.1  \\
KS                   & 0.05  & 3.0  & 44.8  \\
KMR, $R_g=1$         & 0.1   & 8.1  & 510.5  \\
\hline
\end{tabular}
\end{center}
\end{table}

\end{document}